# Excited state calculations using variational quantum eigensolver with spin-restricted ansätze and automatically-adjusted constraints

Shigeki Gocho,[1,2] Hajime Nakamura,[2,3] Shu Kanno,[2,4] Qi Gao,[2,4] Takao Kobayashi,[2,4] Taichi Inagaki,[1,2] and Miho Hatanaka[1,2,*]

[1]School of Fundamental Science and Technology, Faculty of Science and Technology, Keio University, Yokohama 223-8522, Japan. [2]Quantum Computing Center, Keio University, Yokohama 223-8522, Japan. [3]IBM Quantum – IBM Research Tokyo, Tokyo, 103-8510, Japan. [4]Mitsubishi Chemical Corporation, Science & Innovation Center, Yokohama 227-8502, Japan.

*E-mail: hatanaka@chem.keio.ac.jp

## Abstract

The ground and excited state calculations at key geometries, such as the Frank-Condon (FC) and the conical intersection (CI) geometries, are essential for understanding photophysical properties. To compute these geometries on noisy intermediate-scale quantum devices, we proposed a strategy that combined a chemistry-inspired spin-restricted ansatz and a new excited state calculation method called the variational quantum eigensolver under automatically-adjusted constraints (VQE/AC). Unlike the conventional excited state calculation method, called the variational quantum deflation, the VQE/AC does not require the pre-determination of constraint weights and has the potential to describe smooth potential energy surfaces. To validate this strategy, we performed the excited state calculations at the FC and CI geometries of ethylene and phenol blue at the complete active space self-consistent field (CASSCF) level of theory, and found that the energy errors were at most 2 kcal mol$^{-1}$ even on the ibm_kawasaki device.

## Introduction

Computational chemistry has contributed significantly to a better understanding of the mechanisms of chemical phenomena and rational material design. In particular, with the development of the density functional theory (DFT)[1,2] and time-dependent (TD) DFT methods,[3] computational chemistry has become an indispensable technology in a wide range of fields dealing with catalytic, optical, optoelectronic, magnetic, and biomimetic materials. However, the DFT and TDDFT methods are not appropriate for computing quasi-degenerated systems, in which the static electronic correlation makes a large contribution. To take into account electronic correlations, the full-configuration interaction (FCI) method and multireference (MR) calculation methods[4] such as the complete active space self-consistent field (CASSCF),[5] MR configuration interaction,[5] MR coupled-cluster,[6] MR perturbation theory,[7] and MR combined with DFT methods[8] have been proposed. However, their applications to large molecules, in which large active spaces are required, are too demanding. For instance, polynuclear metal complexes such as the $Fe_7MoS_9$ and $Mn_3CaO_4$ complexes in nitrogenase and photosynthetic photosystem II, respectively, have quasi-degenerate characteristics due to the 3d orbitals of the metals, and their computational analyses by MR calculations are still awaited.[9,10] The MR calculation methods are also indispensable for exploring the potential energy surfaces (PESs) of the excited states, especially near the conical intersection (CI) region, which induces the non-radiative deactivation of optical materials.[11-13]

To solve this problem, quantum chemists have given attention to developing novel methods for performing FCI or MR calculations on quantum computers.[10,14-19] This is because quantum computing can, in principle, reduce the computational time for the FCI in a polynomial compared to the classical devices, which require an exponential computation time.[20-24] However, because the current quantum devices, the so-called noisy intermediate-scale quantum (NISQ) devices, are hamstrung by noisiness and short decoherence times, the focus has been on calculation methods that can run on short quantum circuits.[25-27] In the search for quantum advantage with the NISQ devices, various algorithms, including variational quantum algorithm (VQA),[28] full quantum eigensolver with the approximation using the perturbation theory,[29,30] quantum annealing,[31,32] gaussian boson sampling,[33] analog quantum computation,[34] and digital-analog quantum computation,[35] have been proposed. Especially, the VQA for calculating the ground state and the excited states are called the variational quantum eigensolver (VQE)[36] and variational quantum deflation (VQD),[37,38] respectively. These methods have been applied to PESs for small molecules,[39-41] periodic systems,[42-44] energy profiles for lithium batteries,[45,46] and the excitation energies of organic light-emitting diode (OLED) emitters.[47]

Over the past few years, attention has also been given to methodologies for applying the CASSCF calculation, in which the molecular orbitals are optimized with respect to the wavefunction obtained by the VQA.[48-50] CASSCF calculations using quantum devices have the advantage of handling larger active spaces than those using classical devices, thereby enhancing the interpretative and predictive power of the CASSCF calculations. Moreover, these methods were recently extended to perform state-average (SA) CASSCF calculations to provide a balanced description of all the states involved in a photo-excitation system.[51,52] All these pioneering studies, however, have only validated the theoretical accuracy of CASSCF calculations on an ideal quantum computer, which is far from practical enough to be useful for the current NISQ devices.

To obtain sufficient energy accuracy for CASSCF calculations using NISQ devices,[25] promising error mitigation approaches[53,54] have been proposed. However, these techniques still do not provide sufficient accuracy for investigating the PES using the CASSCF method. For example, in the case of the ground state calculations for Li complexes,[45] a deviation of several *mHa* (3-5 kcal mol$^{-1}$) as well as a large spin contamination were observed even with the error mitigation approach. The situation becomes much more pronounced for the excitation energy calculation of OLED emitter molecules[47] because of the 'cost function' for the excited state calculation (see Descriptions of the VQE and VQD in Results and Discussion). Thus, further improvements in the computational techniques are needed to achieve an accuracy that is approximately one order of magnitude higher than those of the current approaches.

To deal with this issue, in this work, we propose an excited state calculation method, named VQE under automatically-adjusted constraints (VQE/AC), and combined it with an appropriate ansatz that restricts the spin multiplicity.[55,56] The VQE/AC is based on a classical constrained optimization algorithm and does not require the cost function, which could cause an error in the VQD calculation. The spin-restricted ansatz can span the subspace of the target spin state, which could avoid the undesired spin contamination. The advantages of this ansatz are as follows: (1) minimum number of variational parameters to fully span the appropriate symmetry subspace and (2) shorter circuit depth than those of other conventional ansätze. To validate our strategy, we perform the CASSCF calculations for ethylene and phenol blue (4-[4-(dimethylamino)phenyl)imino]-2,5-cyclohexadien-1-one, shown in Figure 1). The phenol blue is a nonfluorescent dye, which shows an ultrafast internal conversion from the excited state to the ground state after photoexcitation, and its optical properties have been investigated by both spectroscopic experiments[57,58] and a theoretical simulation.[59] From the viewpoint of an industrial application, the phenol blue is a primary skeletal structure part of indoanilline dyes, which have been applied to cyan-colored materials in photography and dye diffusion thermal transfer printings. To develop a robust dye, it is very important to locate its CI where the nonradiative decay occurs efficiently. In this paper, we first describe the basic idea of VQE, VQD, VQE/AC, and the spin-restricted ansatz. We then apply our approach to the ground and excited states of ethylene at the Frank–Condon (FC) and CI geometries, and compare it with other methods. We also demonstrate the feasibility of our approach by the excited state calculation of the phenol blue dye using the simulators and the real device called ibm_kawasaki.

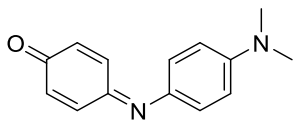

**Fig. 1** Phenol blue

## Results and Discussion

**Descriptions of the VQE and VQD**

The VQE[36] is a ground-state calculation method that uses quantum circuits. The basic idea of the VQE comes from the variational principle: the energy expectation value calculated by any trial wavefunction $\Psi(\boldsymbol{\theta})$ with parameters $\boldsymbol{\theta}$ satisfies the following:

$$\langle\Psi(\boldsymbol{\theta})|\hat{H}|\Psi(\boldsymbol{\theta})\rangle \geq E_0, \qquad (1)$$

where $\hat{H}$ is a given Hamiltonian, and $E_0$ is the minimum eigenvalue. Because this equality is valid only when the trial function is the exact eigenstate of the Hamiltonian (*i.e.*, the wavefunction of the ground state), the energy and wavefunction of the ground state can be obtained by finding the parameters $\boldsymbol{\theta}$ that minimize the energy expectation value. The trial wavefunction in the VQE is constructed using a quantum circuit called *ansatz*, and the energy expectation value is computed *via* quantum measurement. The measurement outcome and parameters are handed over to a classical optimizer, and the parameters are updated so that the energy decreases. The ground state can be obtained by repeating this process until the energy converges.

The excited states can be calculated in a manner similar to the method used by the VQE by minimizing a cost function instead of the energy. This method is called VQD.[37,38] The definition of the cost function depends on the target state, as well as the target system. For instance, the cost function, $C_1(\boldsymbol{\theta})$, for the first excited state can be defined as follows:

$$C_1(\boldsymbol{\theta}) = \langle\Psi(\boldsymbol{\theta})|\hat{H}|\Psi(\boldsymbol{\theta})\rangle + \beta|\langle\Psi(\boldsymbol{\theta})|\Psi_0\rangle|^2, \qquad (2)$$

where $\Psi_0$ is the ground state wavefunction that was previously obtained by the VQE, and $\beta$ is a hyperparameter that must be given before the VQD calculation. The second term of Eq. (2) implies the constraint of searching the subspace orthogonal to the ground state. The parameter $\beta$ needs to be sufficiently large (roughly speaking, greater than the energy difference between the ground state and the excited state).[37,60,61] However, too large $\beta$ could lead to an undesired higher excited state. Another possible cost function, $C_2(\boldsymbol{\theta})$, that can be used to calculate the first singlet excited state as follows:[37]

$$C_2(\boldsymbol{\theta}) = \langle\Psi(\boldsymbol{\theta})|\hat{H}|\Psi(\boldsymbol{\theta})\rangle + \beta|\langle\Psi(\boldsymbol{\theta})|\Psi_0\rangle|^2 + \gamma\langle\Psi(\boldsymbol{\theta})|\hat{S}^2|\Psi(\boldsymbol{\theta})\rangle, \qquad (3)$$

where $\gamma$ is a hyperparameter that constrains the search to a singlet. This cost function is useful for calculating organic molecules whose optical functions are mainly determined by the characteristics of the first singlet excited state ($S_1$) and the singlet ground state ($S_0$). With appropriate hyperparameters, the VQD with the cost function $C_2(\boldsymbol{\theta})$ could give the S$_1$ state, while that with $C_1(\boldsymbol{\theta})$ could give the lowest triplet excited state (T$_1$). When the spin multiplicity of the target state is constrained to be a singlet by the ansatz (as mentioned below), however, the VQD with $C_1(\boldsymbol{\theta})$ could also give the S$_1$ state.

**VQE under automatically-adjusted constraints (VQE/AC)**

Another way to minimize the energy with the constraint of the orthogonality to the ground state is to apply constrained optimization using a linear approximation (COBYLA),[62] which is a numerical optimization method that does not require the derivative of the objective function (*i.e.*, the energy). To obtain the first excited state, the energy expectation value is minimized with the constraint of the overlap such as $|\langle\Psi(\theta)|\Psi_0\rangle|^2 \leq 10^{-4}$. In other words, the weight of the constraint, which corresponds to $\beta$ in VQD, is automatically adjusted within the algorithm of the COBYLA. We named this excited state calculation VQE under automatically-adjusted constraints (VQE/AC). There are two advantages to VQE/AC. First, the cost function tuning is not required, unlike VQD. The second is the applicability to higher excited state calculations because more than two constraints can be considered in the COBYLA. Because the number of constraints does not increase exponentially, the computational cost of a higher excited state calculation should not be too demanding.

**Spin-restricted ansatz**

The spin multiplicity of the trial wavefunction can be restricted using an ansatz called the spin-restricted ansatz. As an example to illustrate the ansatz that restricts the trial wavefunction to a singlet, consider a wavefunction represented by the electronic configurations obtained by the active space with two electrons in two orbitals (*i.e.*, HOMO and LUMO). Under the constraints of the electron number, $N = 2$, and the spin z-projection, $S_z = 0$, the active space can be mapped to the qubit space in the manner of parity mapping[63] as follows:

$$\begin{aligned}
a^\dagger_{\text{HOMO}\uparrow}a^\dagger_{\text{LUMO}\downarrow}|vac\rangle &\rightarrow |11\rangle, \\
a^\dagger_{\text{HOMO}\uparrow}a^\dagger_{\text{HOMO}\downarrow}|vac\rangle &\rightarrow |01\rangle, \\
a^\dagger_{\text{LUMO}\uparrow}a^\dagger_{\text{LUMO}\downarrow}|vac\rangle &\rightarrow |10\rangle, \\
a^\dagger_{\text{LUMO}\uparrow}a^\dagger_{\text{HOMO}\downarrow}|vac\rangle &\rightarrow |00\rangle,
\end{aligned} \quad (4)$$

where $a^\dagger_X$ is the generating operator of an electron in spin orbital X, $|vac\rangle$ is the vacuum state, and the up and down arrows represent two spin eigenstates. Here, the singlet and triplet configurations are represented by a linear combination of Eq. (4). The doubly occupied singlet configurations in the HOMO and LUMO correspond to $|01\rangle$ and $|10\rangle$, respectively. The open-shell singlet and triplet configurations are represented by $(|00\rangle + |11\rangle)/\sqrt{2}$ and $(|00\rangle - |11\rangle)/\sqrt{2}$, respectively. When only the singlet states are focused on, their wavefunctions can be represented by a linear combination of only the singlet configurations. Thus, a quantum circuit that constructs trial functions within the singlet subspace is efficient in avoiding undesired spin contamination. Figure 2 shows a quantum circuit that constructs the singlet subspace. In this circuit, the Pauli X-gate is applied to the second qubit, $q_1$, to prepare the doubly excited configuration, $|10\rangle$, as the initial state. Then, $q_0$ and $q_1$ are transformed into $\sin(\theta_0/2)|01\rangle + \cos(\theta_0/2)|10\rangle$ by the Y-rotation gate, $R_y(\theta_0)$, combined with the CNOT gate. The $R_y(\theta_1)$ and $R_y(-\theta_1)$ pair partly transforms $|01\rangle - |10\rangle$ into $|00\rangle + |11\rangle$ to finally produce a superposition of the three singlet configurations as follows:

$$|\Psi(\boldsymbol{\theta})\rangle = \frac{1}{\sqrt{2}}\left\{\sin(\frac{\theta_0}{2}+\frac{\pi}{4}) - \cos(\frac{\theta_0}{2}+\frac{\pi}{4})\cos\theta_1\right\}|01\rangle$$
$$+ \frac{1}{\sqrt{2}}\left\{\sin(\frac{\theta_0}{2}+\frac{\pi}{4}) + \cos(\frac{\theta_0}{2}+\frac{\pi}{4})\cos\theta_1\right\}|10\rangle \quad (5)$$
$$+ \frac{1}{\sqrt{2}}\cos\left(\frac{\theta_0}{2}+\frac{\pi}{4}\right)\sin\theta_1\,(|00\rangle + |11\rangle).$$

This ansatz is realized by a circuit with minimum gate operations. It should be noted that our spin-restricted ansatz could be expanded for larger active spaces. Gard *et al.*[55] reported ansätze based on the same concept with the Jordan-Wigner mapping[64] and showed the general construction scheme of circuits that enforce particle number and spin for any number of active orbitals and electrons. In section S2 in the SI, we also show the way to construct the spin-restricted ansatz for larger CAS problems (the CAS(4,3) and (4,4) cases as examples) using the parity mapping with two-qubit reduction. Though the number of gates of the spin-restricted ansatz increased as the number of active orbitals increased, the number of parameters to be optimized is still fewer than that of hardware efficient ansätze.

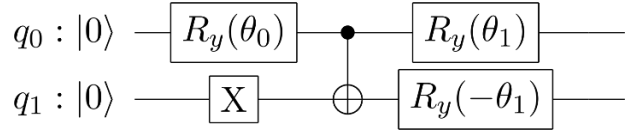

**Fig. 2 Quantum circuit of singlet-restricted ansatz for two electrons in two-orbital system.** X, $R_y$, and the ⊕ connected to a dot represent the Pauli X-gate, Y-rotational gate, and CNOT gate, respectively. $q_0$ and $q_1$ are the labels for the two qubits and the order of tensor products shown in eqs (4) and (5) is $|q_1 q_0\rangle = |q_1\rangle \otimes |q_0\rangle$.

To obtain deeper insights into the singlet subspace, we plotted the energy landscape against the circuit parameters $\theta_0$ and $\theta_1$. Figure 3 shows the energy landscape of ethylene calculated using the CASCI method, whose active space includes two electrons in two orbitals. As shown in Eq. (5), the coefficient of each electronic configuration is represented by the trigonometric functions of parameters $\theta_0$ and $\theta_1$ (in other words, the coefficient changes periodically with respect to $\theta_0$ and $\theta_1$), which results in the periodic energy landscape. In Fig. 3, one of the minimum points, the first-order saddle point, and the second-order saddle point are shown by the white circle, black x, and black triangle, respectively. $S_0$ corresponds to the minimum energy points, which can be determined by minimizing the energy value. $S_1$ corresponds to the first-order saddle point because it is located at the minimum energy point within the subspace that satisfies the orthogonality to $S_0$ (shown by the white solid line in Fig. 3). Therefore, $S_1$ can easily be found using a conventional optimization method under orthogonality constraints. In the same way, the higher (*nth*) singlet excited state, which corresponds to the *nth-order* saddle point, could be obtained by energy minimization within the subspace orthogonal to all the lower singlet states.

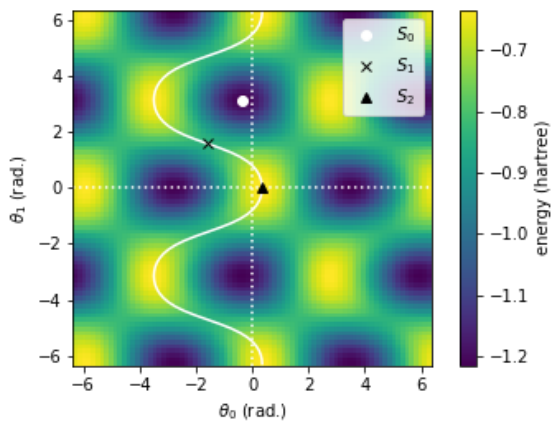

**Fig. 3 Energy landscape (in Hartree) of ethylene plotted against two circuit parameters $\theta_0$ and $\theta_1$ (in radian).** The energies of ethylene were computed at the FC structure using the CASCI/STO-3G method, whose active space included two electrons in two orbitals. One of the minimum energy points ($S_0$), the first-order saddle point ($S_1$), and the second-order saddle point ($S_2$) are shown by the white circle, black x, and black triangle, respectively. The white solid line represents the region where the wavefunction is orthogonal to $S_0$.

**Comparison of ansätze**

Two quantum circuit simulators implemented in Qiskit[65] were used for all the CASSCF calculations. One was the statevector simulator, which simulated the ideal quantum state and did not involve any noise or readout error. The other was a noisy-QASM simulator that used a realistic device (ibmq_belem) noise model. We expect that the appropriate method provides the negligible energy difference between the statevector and noisy-QASM simulators.

First, to examine the dependency on the ansatz, this study focused on the ground state ($S_0$) energy of ethylene at the FC geometry calculated with the state-specific (SS) CASSCF method using two types of ansätze, called the heuristic and chemistry-inspired ansatzes. Figure 4 shows the $S_0$ energy calculated with three heuristic ansatzes, including the real amplitudes (RA) ansatz[66] (with 2 and 6 repetitions (reps) conditions, denoted as RA(2) and RA(6), respectively), the efficient SU2 ansatz,[67] and a chemistry-inspired ansatz, that is, the spin-restricted ansatz. As shown in Fig. 4, when using the statevector simulator, the energy converged to an exact value for all four ansätze. With the noisy-QASM simulator, the calculated energies were higher than the exact value for all the ansätze, but the errors were within 2.5 kcal mol$^{-1}$ at most. It should be noted that the error tended to be larger when using a more complex quantum circuit. As shown in Figs. 2 and S1 (in the SI), the quantum circuit for the spin-restricted ansatz was shorter than those of the heuristic ansätze. In addition, the number of the parameters for the spin-restricted ansatz was only two, which was smaller than the numbers used for the heuristic ansätze (6, 14, and 8 for RA(2), RA(6), and efficient SU2, respectively). It is known that calculations using complex circuits (using many gates and parameters) suffer from the dreaded 'Barren Plateau' of insolvability, where energy minimization becomes difficult due to the flat energy landscape.[68] Thus, the spin-restricted ansatz might have an advantage over the heuristic ansätze by avoiding this problem.

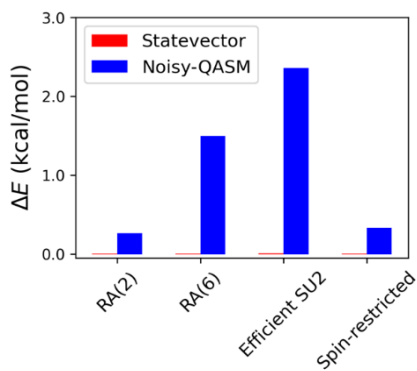

**Fig. 4 Comparison of ansätze for the ground state calculation of ethylene.** The energy deviations $\Delta E$ (in kcal mol$^{-1}$) from the exact values were calculated at the FC geometry using SS-CASSCF with statevector (in red) and noisy-QASM simulators (in blue). Detailed values are shown in Table S1 in Supplemental Information; SI.

Next, we considered the first singlet excited state ($S_1$), as well as the $S_0$ state of ethylene, at the FC and CI geometries. As summarized in Table 1, the calculation methods depend on the ansatz, geometry, and the electronic state. In the case of heuristic ansatz, the calculation methods depend on the geometry. Focusing on the FC geometry, the $S_0$ can be obtained by the VQE, while the $S_1$ can be obtained by the VQD with the cost function $C_2(\boldsymbol{\theta})$ in Eq. (3). The hyperparameter $\beta$, which constrained the search within the subspace orthogonal to $S_0$, was manually adjusted and set to 1. The parameter $\gamma$, which constrained the search within the singlet subspace, needed to be positive and adjusted to 1 because triplet excited states could be more stable than $S_1$. Focusing on the CI geometry, where the $S_0$ and $S_1$ energies were equal, the VQE gave the triplet state ($T_1$) because $T_1$ was always more stable than $S_1$. Therefore, to calculate $S_0$, the VQD with the parameters ($\beta$, $\gamma$) = (0, 1) had to be used instead of the VQE. In the case of the spin-restricted ansatz, on the other hand, the $S_0$ ground state could be obtained by the VQE for any molecular geometry, and the simpler cost function $C_1(\boldsymbol{\theta})$ in Eq. (2) with $\beta = 1$ could be used because the spin multiplicity was constrained to a singlet by the ansatz.

Table 1. Comparison of the calculation methods for $S_0$ and $S_1$.

| Ansatz | geometry | $S_0$ | $S_1$ |
| --- | --- | --- | --- |
| Heuristic | FC | VQE | VQD ($C_2(\boldsymbol{\theta})$, $\beta > 0$) |
| | CI | VQD ($C_2(\boldsymbol{\theta})$, $\beta = 0$) | VQD ($C_2(\boldsymbol{\theta})$, $\beta > 0$) |
| Spin-restricted | Any | VQE | VQD ($C_1(\boldsymbol{\theta})$, $\beta > 0$) |

As shown in Fig. 5, when the statevector simulator was used, every calculation at the FC and CI geometries with any ansatz converged to the exact $S_0$ and $S_1$ energies. However, when using the noisy-QASM simulator, the errors in the $S_0$ and $S_1$ energies differed greatly depending on the ansatz and hyperparameter. Focusing on the energies in Figs. 5(a–c), the errors calculated with the heuristic ansätze were much larger than those found using the spin-restricted ansatz. To understand the reason for the larger errors with the heuristic ansätze, the expected value of spin squared $\langle \hat{S}^2 \rangle$ was focused on (see Table S2 in the SI). The deviation of the spin squared value from the exact value (*i.e.,* zero) was relatively large when the heuristic ansätze were used. Thus, undesired spin contamination could be one of the reasons for the energy errors. In other words, the spin-restricted ansatz has a potential advantage to reduce the error on the energy due to the avoiding undesired subspace, which could be applicable for larger active spaces (see S.2. in the SI). (Note that the errors in the $S_0$ energies calculated by the SA-CASSCF were larger than those calculated by the SS-CASSCF in Fig. 4. It could be understood that inappropriate hyperparameters affected the orbital optimization and eventually both the $S_0$ and $S_1$ energies.) Even though the spin-restricted ansatz was applied, the error in the $S_1$ energy in the CI geometry was as large as 20.96 kcal mol$^{-1}$ (see Fig. 5(d)), while the errors in the $S_1$ energy at the FC as well as the $S_0$ energies were small (up to 0.35 kcal mol$^{-1}$). To clarify the large error in the $S_1$ energy, the coefficients of the three singlet electronic configurations were calculated using Eq. (5). As a result, the major component of the excited state at the CI was the doubly excited electronic configuration; in other words, this calculation converged to $S_2$, not $S_1$. This implied that the exploration of the PESs of the excited states using the VQD could be difficult because the parameter $\beta$ would have to be adjusted for each molecular geometry.

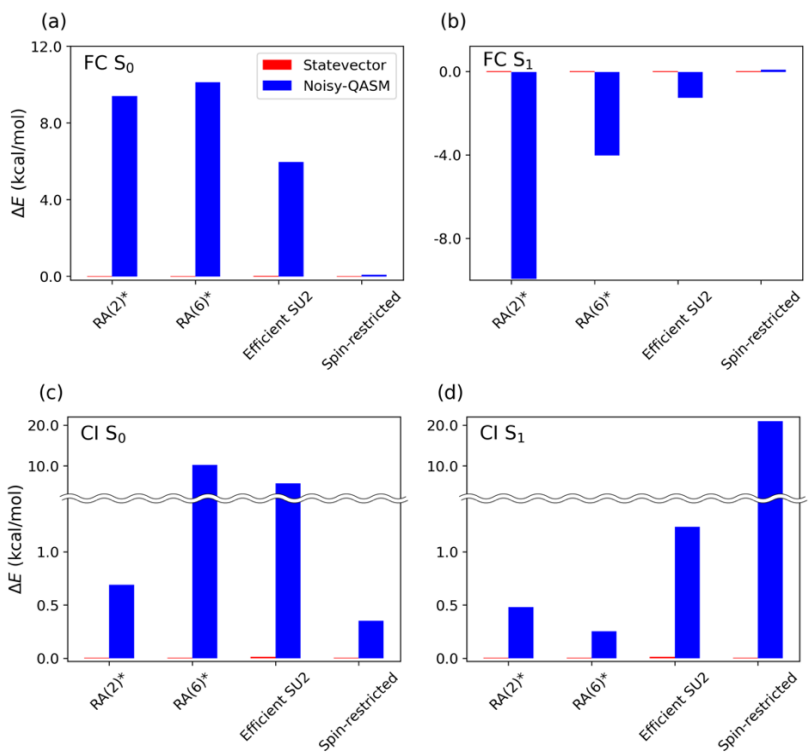

**Fig. 5 Comparison of ansätze for the ground and excited state calculations of ethylene.** Energy deviations $\Delta E$ (in kcal mol$^{-1}$) from the exact values for $S_0$ at the FC (a), $S_1$ at the FC (b), $S_0$ at the CI (c), and $S_1$ at the CI (d) were calculated using the SA-CASSCF method. The errors in energy obtained by the statevector and noisy-QASM simulators are shown in red and blue, respectively. The results labeled with an '*' indicate that the maximum number of orbital rotations was reached. The parameters were $\beta = 1$ for all the ansätze, $\gamma = 1$ for the heuristic ansätze. Detailed values are shown in Table S2 in the SI.

**Comparison of calculation methods for excited states**

Next, we examined the performance of VQD with different $\beta$ and compare them with our proposed excited state calculation method, the VQE/AC. As shown in Fig. 6, we calculated $S_0$ and $S_1$ energies of ethylene with the spin-restricted ansatz and compared these excited state calculation methods. Focusing on the VQD, the $S_1$ energy heavily depended on the parameter $\beta$. When the parameter $\beta$ was set to 1, the excited state at the CI geometry converged to the undesired $S_2$ state, as mentioned above. With the $\beta$ set to 2.5, both $S_0$ and $S_1$ were calculated successfully. When the $\beta$ was larger than 5, even the statevector simulator (and of course the noisy-QASM simulator) gave inaccurate energy values, indicating that these $\beta$ values were not appropriate. Thus, the parameter $\beta$ needed to be carefully and manually adjusted and was 2.5 for ethylene. Although higher excited states such as $S_2$ were beyond the scope of this study, it can be expected that the cost functions could be more difficult to adjust because they must involve constraints on all the lower excited states. On the other hand, when the VQE/AC was applied, the errors in energies at both the FC and CI obtained with the noisy-QASM simulator were very small (up to 0.45 kcal mol$^{-1}$). It should be emphasized that VQE/AC does not require the tuning of the cost function, unlike the VQD. Therefore, the VQE/AC could be used to describe smooth PESs even when using the noisy-QASM simulator, *i.e.*, under realistic device noise.

The VQE/AC could be combined with other ansätze as well as the spin-restricted ansatz. For instance, with the RA(2) ansatz, where six parameters needed to be optimized, the VQE/AC consistently gave relatively small energy deviations (< 2.6 kcal mol$^{-1}$) without any hyperparameter tuning. On the other hand, the VQD parameter $\beta$ affording the smallest energy deviations depended on the molecular geometry (see Table S5). Thus, the VQE/AC could be superior to the VQD for more complicated systems as well. Note that the optimal $\beta$ parameter was also different for each ansatz (see Tables S3 and S5) due to the different energy landscapes along with the $\theta$ parameters.

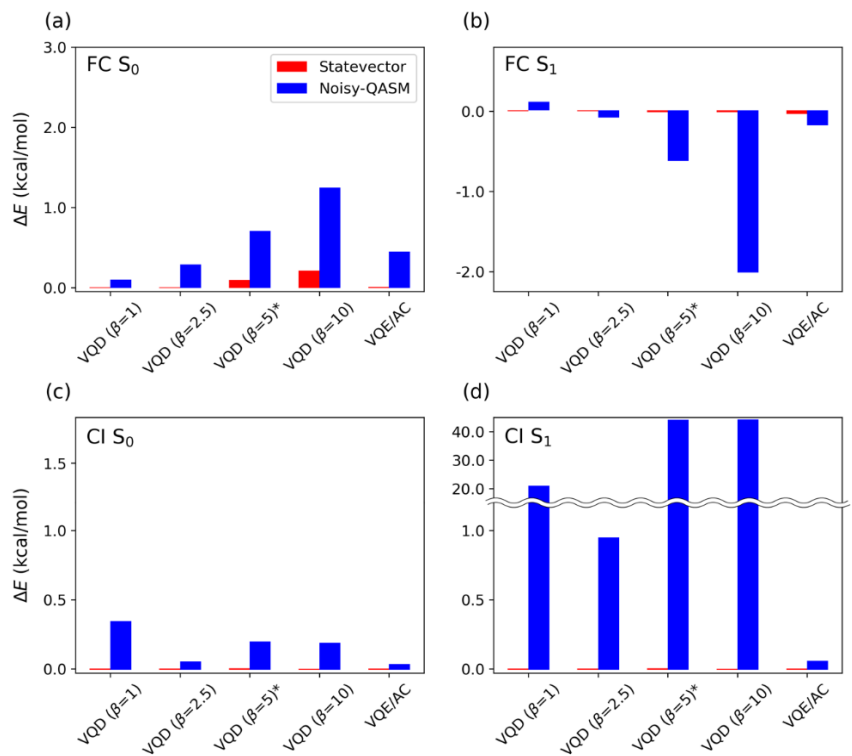

**Fig. 6 Comparison of VQD with different parameter $\beta$ and VQE/AC for ethylene.** Energy deviations $\Delta E$ (in kcal mol$^{-1}$) from the exact values for $S_0$ at the FC (a), $S_1$ at the FC (b), $S_0$ at the CI (c), and $S_1$ at the CI (d) were calculated using the SA-CASSCF method. The errors in energy obtained by the statevector and noisy-QASM simulators are shown in red and blue, respectively. The results labeled with an '*' indicate that the maximum number of orbital rotations was reached. Detailed values are shown in Table S3 in the SI.

**Application to phenol blue**

As previously described, the combination of the spin-restricted ansatz and the VQE/AC enabled to give $S_0$ and $S_1$ energies with an error of less than 1 kcal mol$^{-1}$, even under the realistic device noise model. Next, to verify the applicability of this method to photofunctional molecules, the study focused on a robust dye called phenol blue (see Fig. 7). Focusing on the performance of the VQD method, the most suitable $\beta$ value for both the FC and CI geometries was 1 (in contrast to the value of 2.5 for ethylene), which indicated that the most appropriate value for this parameter heavily depended on the molecule. The errors in energy calculated by the VQD method with parameter $\beta = 1$ were only 0.22 kcal mol$^{-1}$ at most even when using the noisy-QASM simulator. In the case of the VQE/AC, the errors in the $S_0$ and $S_1$ energies were only 0.14 kcal mol$^{-1}$ at most. Therefore, it can be stated that the proposed strategy (VQE/AC with spin-restricted ansatz) is efficient in calculating the excited states, as well as the ground states, of large molecules. This method gave a small error at any geometry without any hyperparameter tuning, which indicated that it is applicable to describe potential energy surfaces of the ground and excited states.

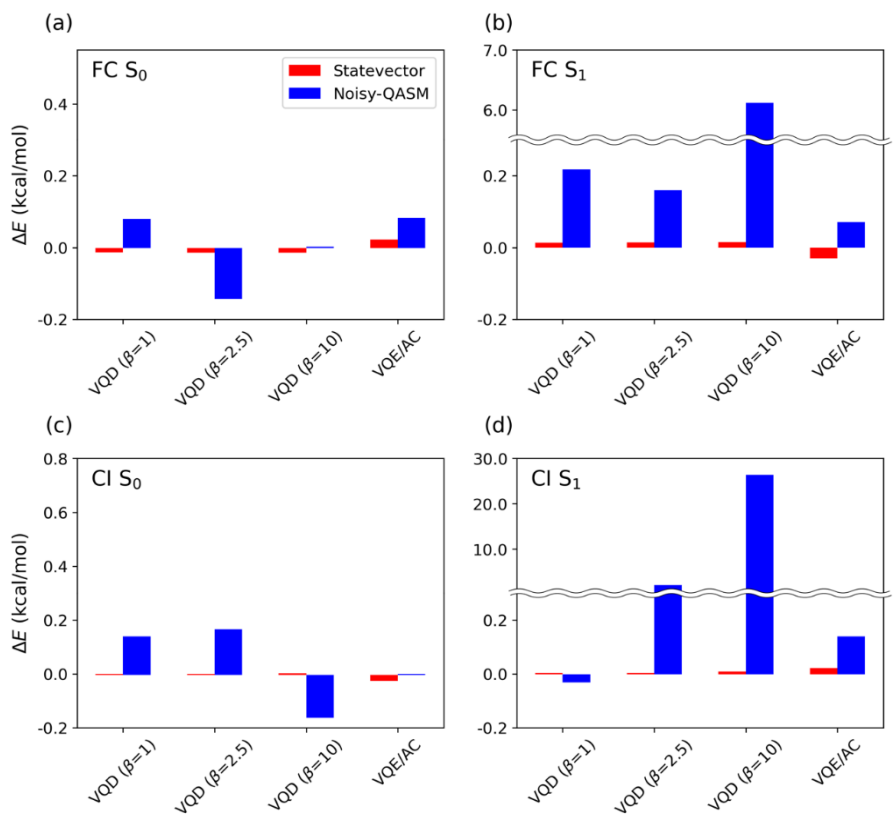

**Fig. 7 Comparison of VQD with different parameter $\beta$ and VQE/AC for phenol blue.** Energy deviations $\Delta E$ (in kcal mol$^{-1}$) from the exact values for $S_0$ at the FC (a), $S_1$ at the FC (b), $S_0$ at the CI (c), and $S_1$ at the CI (d) were calculated using the SA-CASSCF method. The errors in energy obtained by the statevector and noisy-QASM simulators are shown in red and blue, respectively. Detailed values are shown in Table S4 in the SI.

Finally, the ground and excited state energies of phenol blue were measured on the ibm_kawasaki device using the VQE/AC. The energies of the FC and CI geometries were measured twice each as shown in Table 2. All the calculations converged relatively smoothly: the numbers of orbital update iterations were less than 10 in all the calculations. Though the deviations from the exact solutions were larger than those estimated with the noisy-QASM simulator, they were at most 2 kcal mol$^{-1}$ and 0.5 kcal mol$^{-1}$ for the state energies and excitation energies, respectively. It should be noted that the energy deviations at the CI geometry were as small as 0.5 kcal mol$^{-1}$, which were surprisingly small and showed the high potential to achieve the exploration of the CI geometries. The deviations at the FC geometry, on the other hand, were larger than those at the CI. This could be attributed to the fact that the Hamiltonian structure (Pauli string) at the FC geometry was more sensitive to the device noise than that at the CI geometry. Though precise geometry optimization may still be difficult with an error of 2 kcal mol$^{-1}$, it could be improved by developing methodologies of purification and error mitigation as well as hardware.

**Table 2.** Energy deviations $\Delta E$ (in kcal mol$^{-1}$) from the exact values of phenol blue for the $S_0$ and $S_1$ energies at the FC and CI geometries measured on the ibm_kawasaki device.

| Entry | Geometry | $\Delta E$ ($S_0$) | $\Delta E$ ($S_1$) |
|---|---|---|---|
| 1 | FC | 1.68 | 1.64 |
| 2 | FC | 1.82 | 2.03 |
| 3 | CI | 0.37 | 0.04 |
| 4 | CI | 0.49 | 0.01 |

**Conclusions**

This study investigated a ground and excited state calculation method that can tolerate NISQ devices. Two methods were combined, a chemistry-inspired spin-restricted ansatz with parity mapping and an excited-state calculation method, called the VQE/AC method. The advantage of the spin-restricted ansatz was that the wavefunction could be constructed within the subspace of the target spin multiplicity, which reduced the undesired spin contamination. The VQE/AC used a constrained optimization called COBYLA, with the constraint that the overlap integral between the target state and the ground state was smaller than a threshold such as 10$^{-4}$. To validate this strategy, the CASSCF method was used for the singlet ground and excited states of ethylene and phenol blue at the FC and CI geometries. The small errors were obtained in the singlet ground and first excited states (*i.e.*, $S_0$ and $S_1$) on a realistic device noise model (< 0.5 kcal mol$^{-1}$) and the real device 'ibm_kawasaki' (< 2 kcal mol$^{-1}$). The present calculation results are superior to the previous ones using quantum circuits (at least 2-3 kcal mol$^{-1}$).[45,47] Unlike the conventional excited state calculation method called VQD, the VQE/AC does not require any parameter tuning for the cost function. Thus, the VQE/AC could have the advantage of higher excited state calculations (though this was beyond the scope of this study) compared to the VQD. Moreover, it should be emphasized that the ground and excited state energies could be computed with the same calculation condition for any molecular geometry because parameter tuning was not required. Therefore, the VQE/AC could be used to explore PESs of the ground and excited states, even under a realistic device noise model. In other words, the VQE/AC has much potential for achieving geometry optimization of critical structures on and between the ground and excited states using real NISQ devices. Though this study mainly focused on the proof-of-concept demonstration on the real device, the future targets include the photochemistry of large systems, such as biomolecules and polynuclear metal complexes, which require the use of large active spaces to represent their electronic states. According to Ref. 54, the depth of the circuit for the spin-restricted ansatz increased as the number of electronic configurations involved in the active space increased. In addition, the energy error tended to increase with the depth of the circuit. Therefore, the VQE/AC combined with hardware efficient ansätze could be an appropriate strategy to achieve the computation of large systems, which would be the subject of future analyses.

**Methods**

**Workflow and classical computations**

In all the CASSCF calculations,[69,70] the active space included two electrons in two orbitals such as HOMO and LUMO. When only the ground state was focused on, the state-specific (SS) CASSCF was applied. To compute both the ground and first excited states, the state-averaged (SA) CASSCF was applied, in which the average energy of these two states was minimized. The initial (guess) molecular orbitals for the CASSCF were obtained using the Hartree–Fock (HF) method (see Fig. S2 in SI). The basis sets used for ethylene and phenol blue were STO-3G[71] and 6-31G(d),[72] respectively. The molecular geometries of ethylene at the FC and CI were optimized at the same level of theory using the classical CASSCF method (without using the quantum circuit) implemented in the MOLPRO[73-75] and GRRM[76,77] programs. The geometries of phenol blue were obtained from a previous study.[59]

Figure 8 shows the workflow of the CASSCF calculation in this study. As shown in (i) in Fig. 8, we started from calculating the one-body and two-body integrals $h_1$ and $h_2$ (in MO basis) based on the input geometries, spin, and the basis set using the PySCF[78] package. Next, the Qiskit package[65] was used (ii) to prepare the Hamiltonian $\hat{H}$ and the spin-squared operator $\hat{S}^2$ in the second-quantized form and to map them to qubit operators. Then, the VQE for $S_0$ (iii) and VQD for $S_1$ (iv) were conducted, in which the expectation values of the energy (or the cost function) and constraints (for VQE/AC) were measured, and the parameters in the ansatz were updated until the energy/cost function converged. The COBYLA[62] optimizer in the SciPy[79] package was used to update the parameters, and the convergence threshold and the maximum number of iterations were set to $10^{-4}$ atomic units and 100, respectively. When iterations reached the maximum, the result at the last step was taken. Each VQE/VQD was followed by state-tomography (ST) and purification (see below). After the calculations for $S_0$ and $S_1$, (v) the one- and two-particle reduced density matrix (1-RDM and 2-RDM) elements for $S_0$ and $S_1$ were measured using the converged parameters. These RDMs were then averaged with weights of $(S_0, S_1) = (1, 0)$ and $(0.5, 0.5)$ for the SS-CASSCF and SA-CASSCF calculations, respectively. If averaged RDMs and the similarly averaged energy were not converged, the orbitals were updated by modules in the PySCF package and repeated the procedure (ii-v). For the calculations on the simulators, the convergence threshold for the orbitals was set to $10^{-4}$ atomic units for the energy, CI gradients, and orbital rotation gradients. They were altered to $10^{-3}$ atomic units for the energy, $5\times10^{-2}$ atomic units for the gradients in the calculations on the real device.

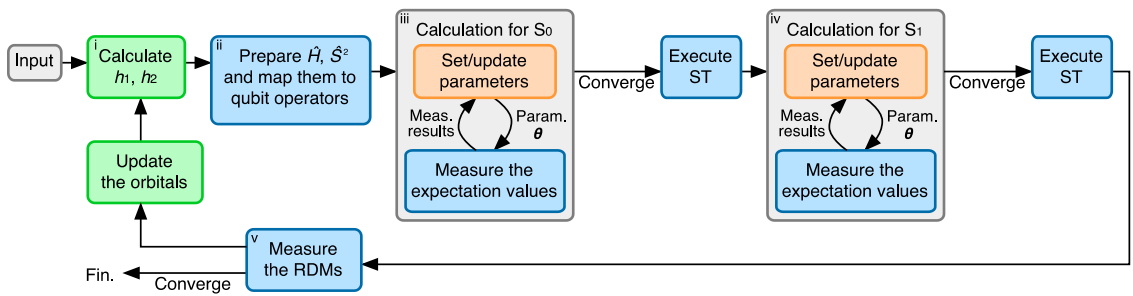

**Fig. 8 Schematic diagram of SA-CASSCF program.** Green, blue and orange boxes indicate that PySCF, Qiskit and SciPy packages were used, respectively.

**Quantum circuits and quantum simulations**

The details of the quantum circuit and measurement were as follows. The parity mapping[60] was used to map the molecular orbitals to qubits. It exploited the symmetry wherein the total electron number and total alpha electron number should be conserved and allowed the qubits to be reduced by two. Therefore, four spin-orbital calculations were conducted on two qubits. The initial parameters were set to $\theta = (0, \pi)$ for the spin-restricted ansatz (which corresponded to the HF state) and all zero for the other ansätzë. Note that the overlap between two states $|\Psi(\theta)\rangle$ and $|\Psi_0\rangle$, $|\langle\Psi(\theta)|\Psi_0\rangle|^2$, was obtained by measuring the quantum circuit of inverted $|\Psi(\theta)\rangle$ combined with $|\Psi_0\rangle$. To measure the expectation values, we used the 'ibm_kawasaki' device and two simulators in the Qiskit[65] package: the statevector simulator, which simulated an ideal quantum state without any noise or readout error, and noisy-QASM simulator, which employed the realistic noise model from 'ibmq_belem' device. For the noisy-QASM simulator and the ibm_kawasaki device, the expectation value was obtained using 8192 shots. The measurement error mitigation implemented in Qiskit was applied for the measurements on the noisy-QASM simulator, otherwise not applied for those on the ibm_kawasaki device because the update of the calibration matrix affected the result. The quantum state-tomography and purification after each VQE/VQD calculation was executed by the following procedure, as found in a previous study[47].

1. Measure density matrix $\rho$.
2. Diagonalize $\rho$ to obtain eigenvalues and eigenvectors with the classical algorithm in SciPy.
3. Assume that the eigenvector $|\psi\rangle$ corresponding to the maximum eigenvalue is the exact state, and re-evaluate the energy as $\langle\psi|\hat{H}|\psi\rangle$.
4. Update parameter set $\theta$ by minimizing $|\langle\psi|\Psi(\theta)\rangle|^2 - 1$.

## Acknowledgments

This work was supported by JSPS KAKENHI Grant no. JP17H06445, 20K05438, and JST Gannt no. JPMJPF2221. We also acknowledge the computer resources provided by the Academic Center for Computing and Media Studies (ACCMS) at Kyoto University and by the Research Center of Computer Science (RCCS) at the Institute for Molecular Science.

**Supporting Information**

**S1. Appendix for the CASSCF(2,2) calculations**

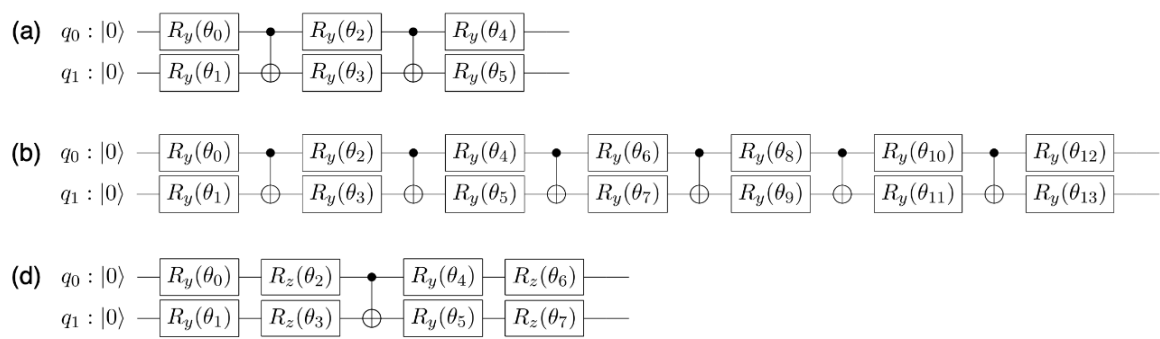

**Fig. S1** The quantum circuits used for the CASSCF calculations of ethylene with the RA ansatz with reps 2 (a), that with reps 6 (b), and the efficient SU(2) ansatz (c).

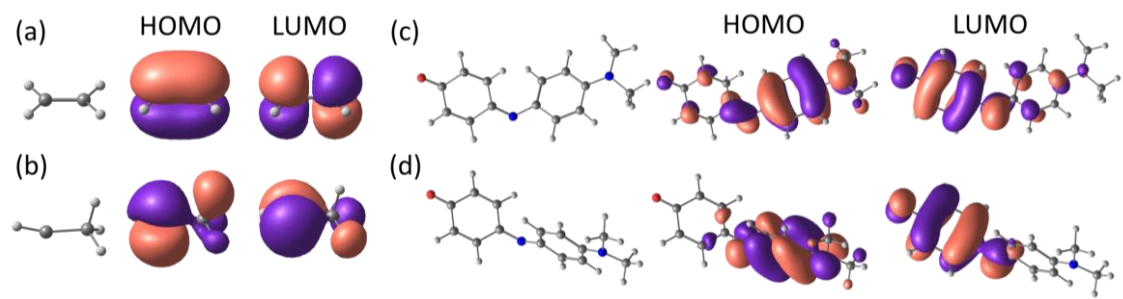

**Fig. S2** The geometries and the active orbitals (HOMO and LUMO) of ethylene at the FC (a) and the CI (b) geometries, and phenol blue at the FC (c) and the CI (d) geometries.

**Table S1.** Comparison of the four ansätze for the ground state calculation. Energy deviations from the exact value $\Delta E$ (in kcal/mol), and spin squared values $S^2$ of the ground state of ethylene at the FC geometry calculated at the SS-CASSCF level of theory.

| Entry | Simulator | Ansatz | $\Delta E$ (kcal/mol) | $S^2$ |
|---|---|---|---|---|
| 1 | Statevector | RA(2) | 0.00 | 0.00 |
| 2 | | RA(6) | 0.00 | 0.00 |
| 3 | | EfficientSU2(1) | 0.01 | 0.00 |
| 4 | | Spin-restricted | 0.00 | 0.00 |
| 5 | Noisy-QASM | RA(2) | 0.27 | 0.01 |
| 6 | | RA(6) | 1.51 | 0.10 |
| 7 | | EfficientSU2(1) | 2.37 | 0.04 |
| 8 | | Spin-restricted | 0.34 | 0.03 |

**Table S2.** Comparison of the four ansätze for the ground and excited state calculations. Energy deviations from the exact value[a] $\Delta E$ (in kcal/mol), and spin squared values $S^2$ of the $S_0$ and $S_1$ states of ethylene at the FC and CI geometries calculated at the SA-CASSCF level of theory.

| Entry | Geometry | Simulator | Ansatz | $\Delta E(S_0)$ (kcal/mol) | $S^2(S_0)$ | $\Delta E(S_1)$ (kcal/mol) | $S^2(S_1)$ |
|---|---|---|---|---|---|---|---|
| 1 | FC | Statevector | RA(2) | 0.00 | 0.00 | 0.00 | 0.00 |
| 2 | | | RA(6) | 0.00 | 0.00 | 0.00 | 0.00 |
| 3 | | | EfficientSU2(1) | 0.01 | 0.00 | 0.00 | 0.00 |
| 4 | | | Spin-restricted | 0.00 | 0.00 | 0.00 | 0.00 |
| 5 | | Noisy-QASM | RA(2) | 9.43 | 0.10 | -9.92 | 0.03 |
| 6 | | | RA(6) | 10.16 | 0.10 | -4.00 | 0.07 |
| 7 | | | EfficientSU2(1) | 5.98 | 0.02 | -1.23 | 0.02 |
| 8 | | | Spin-restricted[b] | 0.11 | 0.02 | 0.11 | 0.01 |
| 9 | | | Spin-restricted[c] | 0.29 | 0.01 | -0.09 | 0.02 |
| 10 | CI | Statevector | RA(2) | 0.00 | 0.00 | 0.00 | 0.00 |
| 11 | | | RA(6) | 0.00 | 0.00 | 0.00 | 0.00 |
| 12 | | | EfficientSU2(1) | 0.00 | 0.00 | 0.00 | 0.00 |
| 13 | | | Spin-restricted | 0.00 | 0.00 | 0.00 | 0.00 |
| 14 | | Noisy-QASM | RA(2) | 0.69 | 0.03 | 0.48 | 0.03 |
| 15 | | | RA(6) | 10.25 | 0.06 | 0.25 | 0.09 |
| 16 | | | EfficientSU2(1) | 5.68 | 0.04 | 1.23 | 0.01 |
| 17 | | | Spin-restricted[b] | 0.35 | 0.02 | 20.96 | 0.01 |
| 18 | | | Spin-restricted[c] | 0.06 | 0.00 | 0.96 | 0.01 |

a) The exact energy differences from $S_0$ at FC were 325.68 kcal/mol and 125.44 kcal/mol for $S_1$ at FC and CI, respectively. b) $\beta = 1$. c) $\beta = 2.5$.

**Table S3.** Comparison of the VQD and VQE/AC with the spin-restricted ansatz. Energy deviations from the exact value[a] $\Delta E$ (in kcal/mol), and spin squared values $S^2$ of the $S_0$ and $S_1$ states of ethylene at the FC and CI geometries calculated at the SA-CASSCF level of theory with the spin-restricted ansatz.

| Entry | Geometry | Simulator | Excited state calculation | $\Delta E(S_0)$ (kcal/mol) | $S^2(S_0)$ | $\Delta E(S_1)$ (kcal/mol) | $S^2(S_1)$ |
|---|---|---|---|---|---|---|---|
| 1 | FC | Statevector | VQD ($\beta = 1$) | 0.00 | 0.00 | 0.00 | 0.00 |
| 2 | | | VQD ($\beta = 2.5$) | 0.00 | 0.00 | 0.00 | 0.00 |
| 3 | | | VQD ($\beta = 5$) | 0.09 | 0.00 | -0.01 | 0.00 |
| 4 | | | VQD ($\beta = 10$) | 0.21 | 0.00 | -0.01 | 0.00 |
| 5 | | | VQE/AC | 0.00 | 0.00 | -0.03 | 0.00 |
| 6 | | Noisy-QASM | VQD ($\beta = 1$) | 0.11 | 0.02 | 0.11 | 0.01 |
| 7 | | | VQD ($\beta = 2.5$) | 0.29 | 0.01 | -0.09 | 0.02 |
| 8 | | | VQD ($\beta = 5$) | 0.71 | 0.02 | -0.63 | 0.02 |
| 9 | | | VQD ($\beta = 10$) | 1.25 | 0.02 | -2.02 | 0.03 |
| 10 | | | VQE/AC | 0.45 | 0.01 | -0.18 | 0.02 |
| 11 | CI | Statevector | VQD ($\beta = 1$) | 0.00 | 0.00 | 0.00 | 0.00 |
| 12 | | | VQD ($\beta = 2.5$) | 0.00 | 0.00 | 0.00 | 0.00 |
| 13 | | | VQD ($\beta = 5$) | 0.00 | 0.00 | 0.01 | 0.00 |
| 14 | | | VQD ($\beta = 10$) | 0.00 | 0.00 | 0.11 | 0.00 |
| 15 | | | VQE/AC | 0.00 | 0.00 | 0.00 | 0.00 |
| 16 | | Noisy-QASM | VQD ($\beta = 1$) | 0.35 | 0.02 | 20.96 | 0.01 |
| 17 | | | VQD ($\beta = 2.5$) | 0.06 | 0.00 | 0.96 | 0.01 |
| 18 | | | VQD ($\beta = 5$) | 0.20 | 0.02 | 44.20 | 0.02 |
| 19 | | | VQD ($\beta = 10$) | 0.19 | 0.01 | 44.32 | 0.01 |
| 20 | | | VQE/AC | 0.04 | 0.01 | 0.06 | 0.02 |

a) The exact energy differences from $S_0$ at FC were 325.68 kcal/mol and 125.44 kcal/mol for $S_1$ at FC and CI, respectively.

**Table S4.** Comparison of the VQD and VQE/AC with the spin-restricted ansatz. Energy deviations from the exact value[a] $\Delta E$ (in kcal/mol), and spin squared values $S^2$ of the $S_0$ and $S_1$ states of phenol blue at the FC and CI geometries calculated at the SA-CASSCF level of theory with the spin-restricted ansatz.

| Entry | Geometry | Simulator | Excited state calculation | $\Delta E(S_0)$ (kcal/mol) | $S^2(S_0)$ | $\Delta E(S_1)$ (kcal/mol) | $S^2(S_1)$ |
|---|---|---|---|---|---|---|---|
| 1 | FC | Statevector | VQD ($\beta = 1$) | -0.01 | 0.00 | 0.01 | 0.00 |
| 2 | | | VQD ($\beta = 2.5$) | -0.01 | 0.00 | 0.01 | 0.00 |
| 3 | | | VQD ($\beta = 10$) | -0.01 | 0.00 | 0.02 | 0.00 |
| 4 | | | VQE/AC | 0.02 | 0.00 | -0.03 | 0.00 |
| 5 | | Noisy-QASM | VQD ($\beta = 1$) | 0.08 | 0.01 | 0.22 | 0.01 |
| 6 | | | VQD ($\beta = 2.5$) | -0.14 | 0.02 | 0.16 | 0.02 |
| 7 | | | VQD ($\beta = 10$) | 0.00 | 0.02 | 6.12 | 0.02 |
| 8 | | | VQE/AC | 0.08 | 0.01 | 0.07 | 0.01 |
| 9 | CI | Statevector | VQD ($\beta = 1$) | 0.00 | 0.00 | 0.00 | 0.00 |
| 10 | | | VQD ($\beta = 2.5$) | 0.00 | 0.00 | 0.00 | 0.00 |
| 11 | | | VQD ($\beta = 10$) | 0.00 | 0.00 | 0.01 | 0.00 |
| 12 | | | VQE/AC | -0.02 | 0.00 | 0.02 | 0.00 |
| 13 | | Noisy-QASM | VQD ($\beta = 1$) | 0.14 | 0.02 | -0.03 | 0.03 |
| 14 | | | VQD ($\beta = 2.5$) | 0.17 | 0.02 | 2.05 | 0.03 |
| 15 | | | VQD ($\beta = 10$) | -0.16 | 0.00 | 26.41 | 0.03 |
| 16 | | | VQE/AC | 0.00 | 0.02 | 0.14 | 0.01 |

a) The exact energy differences from $S_0$ at FC were 85.85 kcal/mol and 46.39 kcal/mol for $S_1$ at FC and CI, respectively.

**Table S5.** Comparison of the VQD and VQE/AC with the RA(2) ansatz.[a] Energy deviations from the exact value[b] $\Delta E$ (in kcal/mol) and spin squared values $S^2$ of the $S_0$ and $S_1$ states of ethylene at the FC and CI geometries calculated at the SA-CASSCF level of theory using the noisy-QASM simulator.

| Entry | Geometry | Excited state calculation | $\Delta E(S_0)$ (kcal/mol) | $S^2(S_0)$ | $\Delta E(S_1)$ (kcal/mol) | $S^2(S_1)$ |
|---|---|---|---|---|---|---|
| 1 | FC | VQD ($\beta = 1$) | 9.43 | 0.10 | -9.92 | 0.03 |
| 2 | | VQD ($\beta = 2.5$) | 6.09 | 0.04 | -7.49 | 0.04 |
| 3 | | VQD ($\beta = 5$)[c] | 0.75 | 0.04 | -1.13 | 0.05 |
| 4 | | VQD ($\beta = 10$) | 8.49 | 0.05 | 8.91 | 0.03 |
| 5 | | VQE/AC | 2.60 | 0.06 | -1.45 | 0.04 |
| 6 | CI | VQD ($\beta = 1$)[c] | 0.69 | 0.03 | 0.48 | 0.03 |
| 7 | | VQD ($\beta = 2.5$) | 0.55 | 0.03 | 26.15 | 0.03 |
| 8 | | VQD ($\beta = 5$) | 8.03 | 0.04 | 35.51 | 0.03 |
| 9 | | VQD ($\beta = 10$) | 9.13 | 0.05 | 0.89 | 0.04 |
| 10 | | VQE/AC | -0.02 | 0.04 | 0.08 | 0.04 |

a) To preserve the spin multiplicity, the $S^2$ was constrained to be zero by the COBYLA algorithm for the VQE/AC, and the spin penalty term was added to the cost function with $\gamma = 1$ for the VQD. This spin constraint/penalty was applied expect for the $S_0$ at FC geometry.
b) The exact energy differences from $S_0$ at FC were 325.68 kcal/mol and 125.44 kcal/mol for $S_1$ at FC and CI, respectively.
c) The optimal $\beta$ parameters, which gave the smallest energy deviations, for the FC and CI geometries were 5 and 1, respectively.

## S2. Appendix for the CASSCF(4,3) and CASSCF(4,4) calculations

To demonstrate the scalability of our strategy, we applied the VQE/AC with the spin-restricted ansatz for the CASSCF calculations with larger active spaces. We applied the CASSCF(4,3) calculation for formaldehyde, in which 4 electrons in 3 active orbitals in Fig S3 were included in the active space, and the CASSCF(4,4) calculation for *trans*-butadiene, in which 4 electrons in 4 active orbitals in Fig S4 were included.

### S2.1 Spin-restricted ansätze for the singlet CAS(4,3) and CAS(4,4) systems

To build the spin-restricted ansätze for the CAS(4,3) and (4,4) systems, we listed up all the singlet configuration state functions (CSFs) using the distinct row table (DRT)[S1] and mapped to the qubit space in the manner of parity mapping with two-qubit reduction as shown in Tables S6 and S7. To span the singlet subspace by these CSFs, we focused on two building blocks $G_1$ and $G_2$ shown in Fig S5a.[S2] The operations of the $G_1$ and $G_2$ gates are the Givens rotations within the subspaces of ($|01\rangle$, $|10\rangle$) and ($|0011\rangle$, $|1100\rangle$), respectively (See Fig. S5a for the matrix representations of the $G_1$ and $G_2$ gates).

Based on these gates, we built the spin-restricted ansatz for the singlet CAS(4,3) system as shown in Fig. S5b. The procedure to build the ansatz was as follows. First, the CSFs containing $N$ qubits with the value "0" were classified into group $N$. In other words, the CSFs whose indexes were $w$ = 0, 1, 2, 3, 4, and 5 were categorized into the group $N$ = 2, 1, 2, 0, 1, 2, respectively. Next, the basis sets $|q_3q_2q_1q_0\rangle$ were generated in order from basis with small $N$. Starting from the basis with the smallest group number $N$ = 0 (*i.e.,* $w$ = 3), one of the bases contained in each CSF with $N$ = 1 ($|1110\rangle$ in $w$ = 4 and $|1101\rangle$ in $w$ = 1) and one of the bases with $N$ = 2 ($|1010\rangle$ in $w$ = 5) were generated by the $Ry$, the $G_1$, and the controlled-$Ry$ gates, respectively. Next, the linear combination of bases in $N$ = 1 were made by the $G_1$ gates with the fixed angle ($\pi/2$), denoted as $G_1(\pi/2)$, to form the CSFs $w$ = 1 and 4. Then, the remaining bases with $N$ = 2 were generated from $|1010\rangle$. The basis $|0101\rangle$ ($w$ = 0) were generated by the application of the $G_2$ gate to $|1010\rangle$. The basis $|1001\rangle$ (in $w$ = 2) were generated by the application of the double controlled-$Ry$ and double controlled-X gates to $|1010\rangle$ and then the linear combination for the CSF ($w$ = 2) were formed by the application of the $G_2(\pi/2)$ gate to $|1001\rangle$.

In the case of the singlet CAS(4,4) system, two CSFs $w$ = 9 and 11 included the same functions, such as (i) $|110011\rangle + |011110\rangle$, (ii) $|111010\rangle + |010111\rangle$, and (iii) $|001100\rangle + |100001\rangle$. To construct the circuit making the linear combination between two specific bases, it was easier to think in a backward way, *i.e.*, construct the circuit to convert the linear combination of two bases to one of them. Thus, we set the gates to generate the functions (i), (ii), and (iii) from one of their bases at the end of the circuit. For instance, the circuit generating the function (i) ($|110011\rangle + |011110\rangle$) from a basis $|110011\rangle$ were designed as follows (see the gates highlighted in orange in Fig. S6): we first focused on the two bases $|110011\rangle$ ($b$ = 15) and $|011110\rangle$ ($b$ = 16) in the function (i) and found the four qubits $q_5q_3q_2q_0$, which had "0011" in one basis ($b$ = 15) and "1100" in the other ($b$ = 16). (See Table S8 for the index of each basis $b$.) To convert "0011+1100" to "0011" by applying the $G_2(\pi/2)$ gate in a backward way, we made the pair of "0011" and "1100" on adjacent qubits $q_3q_2q_1q_0$ by the CNOT gate (Gate-1 in Fig. S6). However, other bases in the CSFs $w$ = 4, 7, 12 15 also included "0011"

or "1100" on $q_3q_2q_1q_0$, which caused undesired linear combinations by applying the $G_2(\pi/2)$ gate (shown in red in Table S8). To avoid it, double controlled-X gates (Gate-2, 3, 4, 5) were added to retain "0011" and "1100" only on the bases $b$ = 15 and 16, respectively. And then, the controlled-X gate (Gate-6) was added to equalize the $q_5q_4$ qubits of the target bases $b$ = 15 and 16, which resulted in the conversion from |111100⟩+|110011⟩ to |110011⟩ by the $G_2(\pi/2)$ gate (Gate-7). Next, we focused on the bases $b$ = 17 and 18 included in the function (ii). These bases $b$ = 17 and 18 were converted to |011110⟩ and |010001⟩, respectively, by the seven gates (Gate-$n$; $n$ = 1,···,7). Thus, to find four qubits with "0011" in one and "1100" in the other, we needed to add controlled-X gate (Gate-8) (see $q_3q_1q_2q_0$ in blue in Table S8). Then, we also added the gates to avoid undesired linear combinations, followed by the $G_2(\pi/2)$ gate to convert the two bases ($b$ = 17 and 18) in the function (ii) to one. All the other basis pairs included with equal weights in the CSFs were also converted to one in the same manner above (see the CSF index $w$ on each $G_2(\pi/2)$ gate in Fig. S6.) The function pairs (i, ii) and (i, ii, iii) in the CSFs $w$ = 9 and 11, respectively, were also converted by the $G_2$ gate highlighted in green in Fig. S6. To summarize so far, the gates between the green and orange blocks in Fig. S6 had the role to make the linear combinations in all the CSFs in Table S7 from 20 bases. Thus, we finally added the gates to generate these 20 bases prior to the $G_2(2\cos^{-1}\sqrt{2/3})$ gate. The circuit to generate 20 bases was built in a similar manner to that of the CAS(4,3) system. We started from a basis |000000⟩ (in group $N$ = 6) and generated bases containing a "1" (in group $N$ = 5) by the $Ry$ and $G_1$ gates, then generated bases containing two "1" (in group $N$ = 4) by the controlled-$Ry$ gate, followed by the gates for making other bases.

**Table S6.** All the singlet CSFs in the CAS(4,3) system represented as the linear combination of Slater determinants and as the manner of parity mapping with two-qubit reduction.

| $w$[a] | CSFs as the linear combination of Slater determinants [b] | CSFs in the manner of parity mapping with two-qubit reduction |
|---|---|---|
| 0 | $\|\varphi_1\bar{\varphi}_1\varphi_2\bar{\varphi}_2\|$ | $\|0101\rangle$ |
| 1 | $(\|\varphi_1\bar{\varphi}_1\varphi_2\bar{\varphi}_3\| - \|\varphi_1\bar{\varphi}_1\bar{\varphi}_2\varphi_3\|)/\sqrt{2}$ | $(\|1101\rangle + \|0111\rangle)/\sqrt{2}$ |
| 2 | $(\|\varphi_1\varphi_2\bar{\varphi}_2\bar{\varphi}_3\| - \|\bar{\varphi}_1\varphi_2\bar{\varphi}_2\varphi_3\|)/\sqrt{2}$ | $(\|1001\rangle + \|0110\rangle)/\sqrt{2}$ |
| 3 | $\|\varphi_1\bar{\varphi}_1\varphi_3\bar{\varphi}_3\|$ | $\|1111\rangle$ |
| 4 | $(\|\varphi_1\bar{\varphi}_2\varphi_3\bar{\varphi}_3\| - \|\bar{\varphi}_1\varphi_2\varphi_3\bar{\varphi}_3\|)/\sqrt{2}$ | $(\|1011\rangle + \|1110\rangle)/\sqrt{2}$ |
| 5 | $\|\varphi_2\bar{\varphi}_2\varphi_3\bar{\varphi}_3\|$ | $\|1010\rangle$ |

a) The index of CSF called "the weight of the walk" in the distinct row table (DRT).
b) $\varphi_i$ and $\bar{\varphi}_i$ represent the $i$-th $\alpha$ and $\beta$ spin orbitals, respectively.

**Table S7.** All the singlet CSFs in the CAS(4,4) represented as the linear combination of Slater determinants and as the manner of parity mapping with two-qubit reduction.

| $w^{a)}$ | CSFs as the linear combination of Slater determinants [b)] | CSFs in the manner of parity mapping with two-qubit reduction |
|---|---|---|
| 0 | $\|\varphi_1\bar{\varphi}_1\varphi_2\bar{\varphi}_2\|$ | $\|001001\rangle$ |
| 1 | $(\|\varphi_1\bar{\varphi}_1\varphi_2\bar{\varphi}_3\| - \|\varphi_1\bar{\varphi}_1\bar{\varphi}_2\varphi_3\|)/\sqrt{2}$ | $(\|011001\rangle + \|001011\rangle)/\sqrt{2}$ |
| 2 | $(\|\varphi_1\bar{\varphi}_2\bar{\varphi}_2\bar{\varphi}_3\| - \|\bar{\varphi}_1\varphi_2\bar{\varphi}_2\varphi_3\|)/\sqrt{2}$ | $(\|010001\rangle + \|001010\rangle)/\sqrt{2}$ |
| 3 | $\|\varphi_1\bar{\varphi}_1\varphi_3\bar{\varphi}_3\|$ | $\|011011\rangle$ |
| 4 | $(\|\varphi_1\bar{\varphi}_2\varphi_3\bar{\varphi}_3\| - \|\bar{\varphi}_1\varphi_2\varphi_3\bar{\varphi}_3\|)/\sqrt{2}$ | $(\|010011\rangle + \|011010\rangle)/\sqrt{2}$ |
| 5 | $\|\varphi_2\bar{\varphi}_2\varphi_3\bar{\varphi}_3\|$ | $\|010010\rangle$ |
| 6 | $(\|\varphi_1\bar{\varphi}_1\varphi_2\bar{\varphi}_4\| - \|\varphi_1\bar{\varphi}_1\bar{\varphi}_2\varphi_4\|)/\sqrt{2}$ | $(\|111001\rangle + \|001111\rangle)/\sqrt{2}$ |
| 7 | $(\|\varphi_1\varphi_2\bar{\varphi}_2\bar{\varphi}_4\| - \|\bar{\varphi}_1\varphi_2\bar{\varphi}_2\varphi_4\|)/\sqrt{2}$ | $(\|110001\rangle + \|001110\rangle)/\sqrt{2}$ |
| 8 | $(\|\varphi_1\bar{\varphi}_1\varphi_3\bar{\varphi}_4\| - \|\varphi_1\bar{\varphi}_1\bar{\varphi}_3\varphi_4\|)/\sqrt{2}$ | $(\|111011\rangle + \|011111\rangle)/\sqrt{2}$ |
| 9 | $(\|\varphi_1\bar{\varphi}_2\varphi_3\bar{\varphi}_4\| + \|\bar{\varphi}_1\varphi_2\bar{\varphi}_3\varphi_4\|$ $-\|\bar{\varphi}_1\varphi_2\varphi_3\bar{\varphi}_4\| - \|\varphi_1\bar{\varphi}_2\bar{\varphi}_3\varphi_4\|)/2$ | $(\|110011\rangle + \|011110\rangle$ $+ \|111010\rangle + \|010111\rangle)/2$ |
| 10 | $(\|\varphi_2\bar{\varphi}_2\varphi_3\bar{\varphi}_4\| - \|\varphi_2\bar{\varphi}_2\bar{\varphi}_3\varphi_4\|)/\sqrt{2}$ | $(\|110010\rangle + \|010110\rangle)/\sqrt{2}$ |
| 11 | $(-\|\varphi_1\bar{\varphi}_2\varphi_3\bar{\varphi}_4\| - \|\bar{\varphi}_1\varphi_2\bar{\varphi}_3\varphi_4\|$ $-\|\bar{\varphi}_1\varphi_2\varphi_3\bar{\varphi}_4\| - \|\varphi_1\bar{\varphi}_2\bar{\varphi}_3\varphi_4\|$ $+2\|\bar{\varphi}_1\bar{\varphi}_2\varphi_3\varphi_4\| + 2\|\varphi_1\varphi_2\bar{\varphi}_3\bar{\varphi}_4\|)/2\sqrt{3}$ | $(\|110011\rangle + \|011110\rangle$ $-\|111010\rangle - \|010111\rangle$ $+2\|001100\rangle + 2\|100001\rangle)/2\sqrt{3}$ |
| 12 | $(\|\varphi_1\varphi_3\bar{\varphi}_3\bar{\varphi}_4\| - \|\bar{\varphi}_1\varphi_3\bar{\varphi}_3\varphi_4\|)/\sqrt{2}$ | $(\|100011\rangle + \|011100\rangle)/\sqrt{2}$ |
| 13 | $(\|\varphi_2\varphi_3\bar{\varphi}_3\bar{\varphi}_4\| - \|\bar{\varphi}_2\varphi_3\bar{\varphi}_3\varphi_4\|)/\sqrt{2}$ | $(\|100010\rangle + \|010100\rangle)/\sqrt{2}$ |
| 14 | $\|\varphi_1\bar{\varphi}_1\varphi_4\bar{\varphi}_4\|$ | $\|111111\rangle$ |
| 15 | $(\|\varphi_1\bar{\varphi}_2\varphi_4\bar{\varphi}_4\| - \|\bar{\varphi}_1\varphi_2\varphi_4\bar{\varphi}_4\|)/\sqrt{2}$ | $(\|110111\rangle + \|111110\rangle)/\sqrt{2}$ |
| 16 | $\|\varphi_2\bar{\varphi}_2\varphi_4\bar{\varphi}_4\|$ | $\|110110\rangle$ |
| 17 | $(\|\varphi_1\bar{\varphi}_3\varphi_4\bar{\varphi}_4\| - \|\bar{\varphi}_1\varphi_3\varphi_4\bar{\varphi}_4\|)/\sqrt{2}$ | $(\|100111\rangle + \|111100\rangle)/\sqrt{2}$ |
| 18 | $(\|\varphi_2\bar{\varphi}_3\varphi_4\bar{\varphi}_4\| - \|\bar{\varphi}_2\varphi_3\varphi_4\bar{\varphi}_4\|)/\sqrt{2}$ | $(\|100110\rangle + \|110100\rangle)/\sqrt{2}$ |
| 19 | $\|\varphi_3\bar{\varphi}_3\varphi_4\bar{\varphi}_4\|$ | $\|100100\rangle$ |

a) The index of CSF called "weight of walk" in the distinct row table (DRT).
b) $\varphi_i$ and $\bar{\varphi}_i$ represent the $i$-th $\alpha$ and $\beta$ spin orbitals, respectively.

**Table S8.** The list of bases ($b$)[a]) included in the singlet CSFs ($w$)[b]) and their change log by the application of each gate shown in Fig. S6.

| $w$[b] | $b$[a] | Bases in CSFs | Gate-1 [c] cX | Gate-2 [c] ccX | Gate-3 [c] ccX | Gate-4 [c] ccX | Gate-5 [c] ccX | Gate-6 [c] cX | Gate-7 $G_2(\pi/2)$ | Gate-8 [c] cX |
|---|---|---|---|---|---|---|---|---|---|---|
| 0 | 0 | $\|001001\rangle$ | $\|001001\rangle$ | | | $\|001101\rangle$ | | $\|101101\rangle$ | | |
| 1 | 1 | $\|011001\rangle$ | $\|011001\rangle$ | | | $\|011001\rangle$ | | $\|111001\rangle$ | | |
|   | 2 | $\|001011\rangle$ | $\|001011\rangle$ | | | $\|001111\rangle$ | | $\|101111\rangle$ | | |
| 2 | 3 | $\|010001\rangle$ | $\|010001\rangle$ | $\|010101\rangle$ | | $\|010101\rangle$ | | $\|010101\rangle$ | | $\|010001\rangle$ |
|   | 4 | $\|001010\rangle$ | $\|001010\rangle$ | $\|001010\rangle$ | | $\|001110\rangle$ | | $\|101110\rangle$ | | $\|101110\rangle$ |
| 3 | 5 | $\|011011\rangle$ | $\|011011\rangle$ | | | | | $\|111011\rangle$ | | |
| 4 | 6 | $\|010011\rangle$ | $\|010011\rangle$ | $\|010111\rangle$ | | | | $\|010111\rangle$ | | $\|010011\rangle$ |
|   | 7 | $\|011010\rangle$ | $\|011010\rangle$ | $\|011010\rangle$ | | | | $\|111010\rangle$ | | $\|111010\rangle$ |
| 5 | 8 | $\|010010\rangle$ | $\|010010\rangle$ | $\|010110\rangle$ | | | | $\|010110\rangle$ | | $\|010010\rangle$ |
| 6 | 9 | $\|111001\rangle$ | $\|111001\rangle$ | | $\|111101\rangle$ | $\|111101\rangle$ | | $\|011101\rangle$ | | $\|011001\rangle$ |
|   | 10 | $\|001111\rangle$ | $\|001101\rangle$ | | $\|001101\rangle$ | $\|001001\rangle$ | | $\|101001\rangle$ | | $\|101001\rangle$ |
| 7 | 11 | $\|110001\rangle$ | $\|110001\rangle$ | | | $\|110001\rangle$ | | $\|110001\rangle$ | | |
|   | 12 | $\|001110\rangle$ | $\|001110\rangle$ | | | $\|001000\rangle$ | | $\|101000\rangle$ | | |
| 8 | 13 | $\|111011\rangle$ | $\|111011\rangle$ | | $\|111111\rangle$ | | | $\|011111\rangle$ | | $\|011011\rangle$ |
|   | 14 | $\|011111\rangle$ | $\|011101\rangle$ | | $\|011101\rangle$ | | | $\|111101\rangle$ | | $\|111101\rangle$ |
| 9 | 15 | $\|110011\rangle$ | $\|110011\rangle$ | $\|110011\rangle$ | $\|110011\rangle$ | | | $\|110011\rangle$ | $\|110011\rangle$ | $\|110011\rangle$ |
|   | 16 | $\|011110\rangle$ | $\|011100\rangle$ | $\|011100\rangle$ | $\|011100\rangle$ | | | $\|111100\rangle$ | – | – |
|   | 17 | $\|111010\rangle$ | $\|111010\rangle$ | $\|111010\rangle$ | $\|111110\rangle$ | | | $\|011110\rangle$ | $\|011110\rangle$ | $\|011010\rangle$ |
|   | 18 | $\|010111\rangle$ | $\|010101\rangle$ | $\|010001\rangle$ | $\|010001\rangle$ | | | $\|010001\rangle$ | $\|010001\rangle$ | $\|010101\rangle$ |
| 10 | 19 | $\|110010\rangle$ | $\|110010\rangle$ | $\|110010\rangle$ | | | | $\|110010\rangle$ | | $\|110010\rangle$ |
|    | 20 | $\|010110\rangle$ | $\|010100\rangle$ | $\|010000\rangle$ | | | | $\|010000\rangle$ | | $\|010100\rangle$ |
| 11 | 15 | $\|110011\rangle$ | $\|110011\rangle$ | $\|110011\rangle$ | $\|110011\rangle$ | $\|110011\rangle$ | $\|110011\rangle$ | $\|110011\rangle$ | $\|110011\rangle$ | $\|110011\rangle$ |
|    | 16 | $\|011110\rangle$ | $\|011100\rangle$ | $\|011100\rangle$ | $\|011100\rangle$ | $\|011100\rangle$ | $\|011100\rangle$ | $\|111100\rangle$ | – | – |
|    | 17 | $\|111010\rangle$ | $\|111010\rangle$ | $\|111010\rangle$ | $\|111110\rangle$ | $\|111110\rangle$ | $\|111110\rangle$ | $\|011110\rangle$ | $\|011110\rangle$ | $\|011010\rangle$ |
|    | 18 | $\|010111\rangle$ | $\|010101\rangle$ | $\|010001\rangle$ | $\|010001\rangle$ | $\|010001\rangle$ | $\|010001\rangle$ | $\|010001\rangle$ | $\|010001\rangle$ | $\|010101\rangle$ |
|    | 21 | $\|001100\rangle$ | $\|001110\rangle$ | $\|001110\rangle$ | $\|001110\rangle$ | $\|001010\rangle$ | $\|001010\rangle$ | $\|101010\rangle$ | $\|101010\rangle$ | $\|101010\rangle$ |
|    | 22 | $\|100001\rangle$ | $\|100001\rangle$ | $\|100001\rangle$ | $\|100001\rangle$ | $\|100001\rangle$ | $\|100101\rangle$ | $\|100101\rangle$ | $\|100101\rangle$ | $\|100101\rangle$ |
| 12 | 23 | $\|100011\rangle$ | $\|100011\rangle$ | | | | $\|100111\rangle$ | $\|100111\rangle$ | | |
|    | 24 | $\|011100\rangle$ | $\|011110\rangle$ | | | | $\|011110\rangle$ | $\|111110\rangle$ | | |
| 13 | 25 | $\|100010\rangle$ | $\|100010\rangle$ | $\|100010\rangle$ | | | $\|100110\rangle$ | $\|100110\rangle$ | | $\|100110\rangle$ |
|    | 26 | $\|010100\rangle$ | $\|010110\rangle$ | $\|010010\rangle$ | | | $\|010010\rangle$ | $\|010010\rangle$ | | $\|010110\rangle$ |
| 14 | 27 | $\|111111\rangle$ | $\|111101\rangle$ | | $\|111001\rangle$ | | | $\|011001\rangle$ | | $\|011101\rangle$ |
| 15 | 28 | $\|110111\rangle$ | $\|110101\rangle$ | | $\|110101\rangle$ | | | $\|110101\rangle$ | | $\|110101\rangle$ |
|    | 29 | $\|111110\rangle$ | $\|111110\rangle$ | | $\|111000\rangle$ | | | $\|011000\rangle$ | | $\|011100\rangle$ |
| 16 | 30 | $\|110110\rangle$ | $\|110100\rangle$ | | | | | $\|110100\rangle$ | | |
| 17 | 31 | $\|100111\rangle$ | $\|100101\rangle$ | | $\|100101\rangle$ | | $\|100001\rangle$ | $\|100001\rangle$ | | $\|100001\rangle$ |
|    | 32 | $\|111100\rangle$ | $\|111110\rangle$ | | $\|111010\rangle$ | | $\|111010\rangle$ | $\|011010\rangle$ | | $\|011110\rangle$ |
| 18 | 33 | $\|100110\rangle$ | $\|100100\rangle$ | | | | $\|100000\rangle$ | $\|100000\rangle$ | | |
|    | 34 | $\|110100\rangle$ | $\|110110\rangle$ | | | | $\|110110\rangle$ | $\|110110\rangle$ | | |
| 19 | 35 | $\|100100\rangle$ | $\|100110\rangle$ | | | | $\|100010\rangle$ | $\|100010\rangle$ | | |

a) The index of the basis.

b) The index of the CSF called "weight of walk" in the distinct row table (DRT).

c) The names of the gates (Gate-$n$; $n = 1,\cdots, 8$) are shown in Fig. S6. The control and target qubits for each gate are highlighted in green and pink, respectively. The bases that are not changed by each gate operation are omitted.

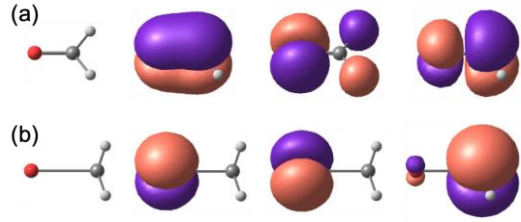

**Fig. S3** The geometries and the three active orbitals of formaldehyde at the FC (a) and the CI (b) geometries obtained by the SA-CASSCF(4,3) calculations with the STO-3G basis set.

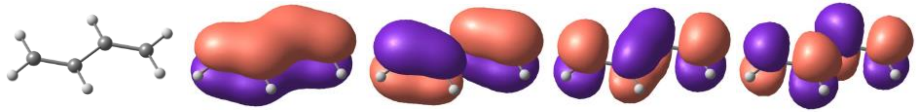

**Fig. S4** The geometries and the four active orbitals of *trans*-butadiene at the FC geometry obtained by the SA-CASSCF(4,4) calculations with the STO-3G basis set.

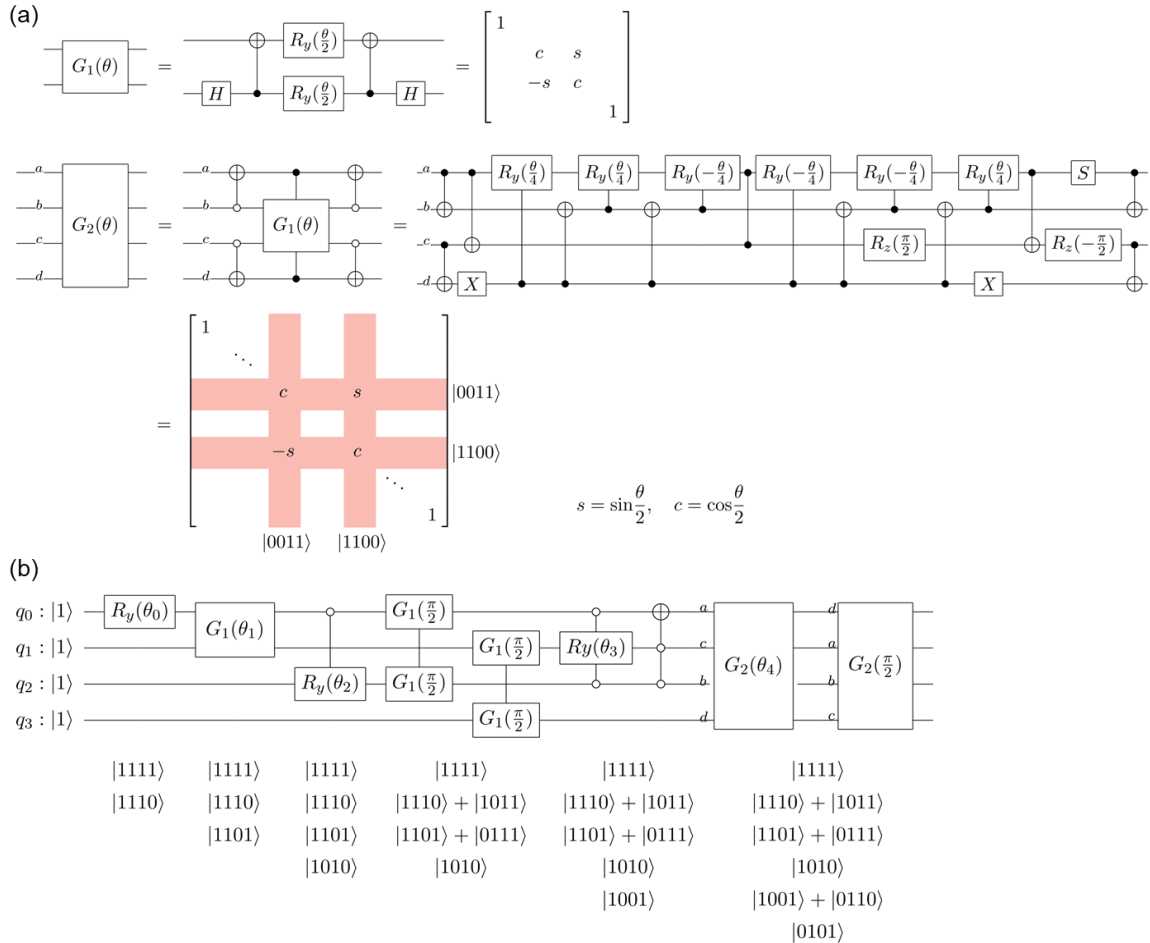

**Fig. S5** The $G_1$ and $G_2$ gates and their matrix representations (a). The quantum circuit of the spin-restricted ansatz for the singlet CAS(4,3) system and the list of the bases generated at each gate (b). $q_0$, $q_1$, $q_2$, and $q_3$ are the labels for the four qubits and the order of tensor products is $|q_3 q_2 q_1 q_0\rangle = |q_3\rangle \otimes |q_2\rangle \otimes |q_1\rangle \otimes |q_0\rangle$.

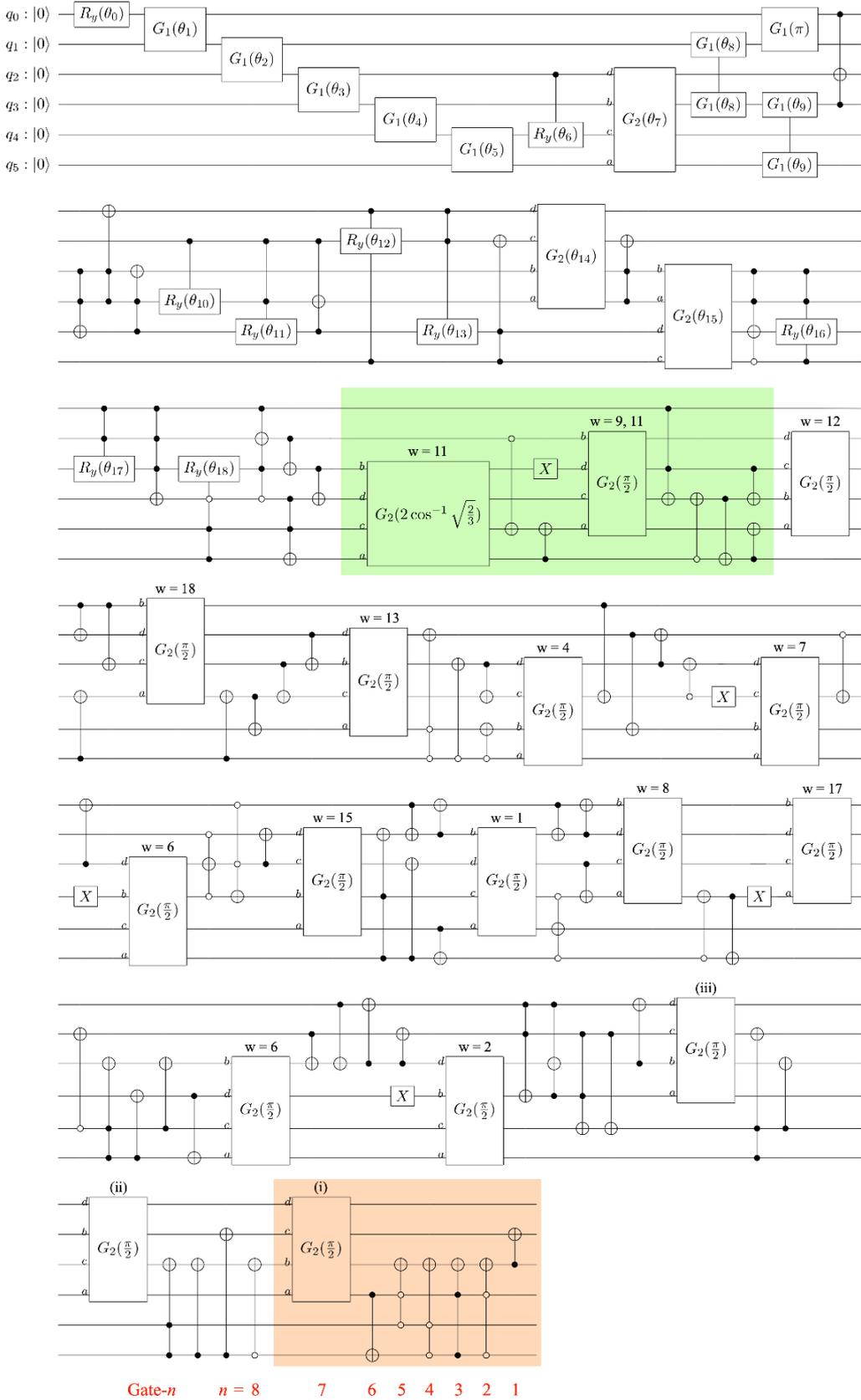

**Fig. S6** The quantum circuit of the spin-restricted ansatz for the singlet CAS(4,4) system. $q_i$ ($i$ = 0-5) are the labels for the six qubits and the order of tensor products is $|q_5q_4q_3q_2q_1q_0\rangle = |q_5\rangle \otimes |q_4\rangle \otimes |q_3\rangle \otimes |q_2\rangle \otimes |q_1\rangle \otimes |q_0\rangle$. The gates generating the function (i) from the basis $b$ = 15 and those generating the CSFs $w$ = 9 and 11 from the functions (i, ii, iii) are highlighted in orange and green, respectively. The linear combinations of specific bases in the CSFs ($w$) and the functions (i, ii, iii) are formed at the $G_2$ gates whose labels are described above. The names of gates (Gate-$n$; $n$ =1-8) correspond to Table S8.

### S2.2 Applications to formaldehyde and *trans*-butadiene

We performed the SA-CASSCF(4,3)/STO-3G calculations of the $S_0$ and $S_1$ states of formaldehyde with different ansätze and simulators as shown in Table S6. First, we examined the reps of the RA ansatz with the statevector simulator at the FC geometry (entries 1-3) and confirmed that the minimal reps required to reproduce the $S_0$ and $S_1$ energies was four (see its quantum circuit in Fig. S7). Comparing the spin-restricted and the RA(4) ansätze, the numbers of the parameters to be optimized were 5 and 20, respectively. The error of the spin-restricted ansatz in the QASM simulator with 8192 shots was smaller than that of the RA(4) ansatz (entries 5 and 6). This could attribute to the fewer parameters and smaller spin contaminations for the spin-restricted ansatz. The same trend was observed for the calculations at the CI geometry (entries 9 and 10). The SA-CASSCF(4,4)/STO-3G calculation of *trans*-butadiene was also examined. As shown in Table S7, the errors in the QASM simulator were larger than those of formaldehyde, however, they were as small as 5 kcal/mol. It should be noted that the spin-restricted ansatz could describe the Hilbert space with the minimum number of parameters. Currently, the number of gates in the spin-restricted ansätze with a large active space is too demanding for the real devices. However, it could be beneficial to have fewer parameters to be optimized when the gate fidelity and the coherent time in the real devices are improved.[S3]

**Table S6.** Comparison of the RA and spin-restricted ansätze for the CAS(4,3) system. Energy deviations from the exact values $\Delta E$ (in kcal/mol) and spin squared values $S^2$ of the $S_0$ and $S_1$ states of formaldehyde at the FC and CI geometries calculated at the SA-CASSCF(4,3)/STO-3G level of theory.

| Entry | Geometry | Simulator | Ansatz | $\Delta E(S_0)$ (kcal/mol) | $S^2(S_0)$ | $\Delta E(S_1)$ (kcal/mol) | $S^2(S_1)$ |
|---|---|---|---|---|---|---|---|
| 1 | FC | Statevector | RA(2) | 0.48 | 0.00 | 0.12 | 0.00 |
| 2 | | | RA(3) | 0.14 | 0.00 | 0.00 | 0.00 |
| 3 | | | RA(4) | 0.01 | 0.00 | 0.00 | 0.00 |
| 4 | | | Spin-restricted | 0.00 | 0.00 | 0.00 | 0.00 |
| 5 | | QASM a) | RA(4) | 2.90 | 0.02 | 19.86 | 0.02 |
| 6 | | | Spin-restricted | 0.78 | 0.00 | 0.83 | 0.01 |
| 7 | CI | Statevector | RA(4) | 0.00 | 0.00 | 0.29 | 0.00 |
| 8 | | | Spin-restricted | 0.00 | 0.00 | 0.00 | 0.00 |
| 9 | | QASM a) | RA(4) | 3.84 | 0.02 | 0.81 | 0.01 |
| 10 | | | Spin-restricted | 1.22 | 0.00 | -0.57 | 0.00 |

a) QASM simulator considering the sampling error with 8192 shots (without noise).

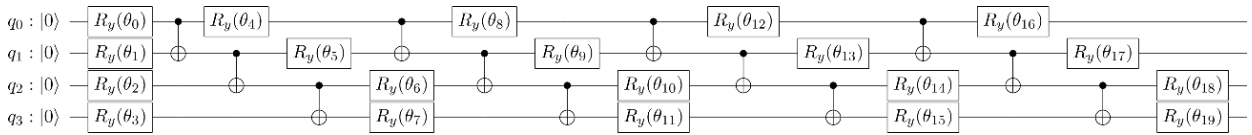

**Fig. S7** The quantum circuit used for the CASSCF(4,3) calculations with the RA ansatz with reps 4.

**Table S7.** The energy deviations $\Delta E$ (in kcal/mol) and spin squared values $S^2$ of the $S_0$ and $S_1$ states of *trans*-butadiene at the FC geometry computed at the SA-CASSCF(4,4)/STO-3G level of theory with the spin-restricted ansatz.

| Entry | Simulator | $\Delta E(S_0)$ (kcal/mol) | $S^2(S_0)$ | $\Delta E(S_1)$ (kcal/mol) | $S^2(S_1)$ |
|---|---|---|---|---|---|
| 1 | Statevector | 0.00 | 0.00 | 0.00 | 0.00 |
| 2 | QASM [a] | 5.05 | -0.01 | 4.60 | 0.02 |

[a] QASM simulator considering the sampling error with 8192 shots (without noise).

## S3. Cartesian coordinates

Ethylene at the FC geometry
```
C    -0.000000000000    0.000000000000   -0.005302010370
C     0.000000000000    0.000000000000    1.335460010375
H     0.000000000000    0.918153736897   -0.576538208909
H    -0.000000000000   -0.918153736897   -0.576538208909
H    -0.000000000000    0.918153736899    1.906696208906
H     0.000000000000   -0.918153736899    1.906696208906
```
Ethylene at the CI geometry
```
C     0.091064191272    0.667402904949   -0.094755763730
C     0.039586415826   -0.340260623235    0.977974481968
H    -0.129316494615    1.485745200663   -0.738632568502
H    -0.025211751438   -1.366684861267    0.611221080196
H    -0.863822866325   -0.154618157022    1.570520406104
H     0.887700505272   -0.291584464087    1.664146363956
```
Phenol Blue at the FC geometry
```
C    3.6568705674    0.9142049564   -0.2228062135
C    2.4361141625    1.4327917986   -0.2624405861
C    1.2242043354    0.6274016737   -0.0281090742
C    1.4374069957   -0.7708496225    0.4010925142
C    2.6553075929   -1.2920406650    0.4684105391
C    3.8692737038   -0.5104528359    0.1159125330
C   -1.1456482157    0.6354807233   -0.1508235139
C   -1.4671488696   -0.6086921252   -0.6870033284
C   -2.7630016884   -1.0874203326   -0.6703626937
C   -3.8063889191   -0.3513954425   -0.0954487138
C   -3.4843419944    0.9201669255    0.4049407161
C   -2.1955695710    1.4012589180    0.3529633989
C   -5.4282407592   -2.0337651063   -0.7913017909
C   -6.1841675019    0.0665065151    0.2491168666
N    0.1139193149    1.2271888083   -0.1748515912
N   -5.0988891798   -0.8510997418   -0.0259600584
O    4.9579204073   -1.0083842205    0.1351222815
H    4.5376431524    1.4944334412   -0.4260204246
H    2.2642523661    2.4641903467   -0.5079519903
H    0.5848390908   -1.3444473259    0.7066926348
H    2.8268843104   -2.2978237767    0.8053209742
H   -0.7135227219   -1.1948445127   -1.1792928178
H   -1.9787740295    2.3865181192    0.7228667665
H   -4.2426795808    1.5493794189    0.8248369078
H   -2.9513687782   -2.0384650612   -1.1257542437
H   -7.1145389852   -0.4825874278    0.2602069189
H   -6.0692643179    0.5203646714    1.2255656735
H   -6.2650721123    0.8612357918   -0.4918349828
```

```
H   -6.4580932017  -2.3000392561  -0.6019744920
H   -5.3030745656  -1.8943896563  -1.8650022206
H   -4.8183660102  -2.8736279978  -0.4825889894
```
Phenol Blue at the CI geometry
```
C    3.2027101937   0.5556864584  -1.0534794716
C    1.9602791158   0.9487822424  -0.7568810868
C    1.2595451126   0.4357384478   0.4083584491
C    1.9775755382  -0.4797814966   1.2791646057
C    3.2200951661  -0.8825314152   0.9962059205
C    3.9357619755  -0.4004677776  -0.2010973471
C   -1.1802342713   0.4570369208   0.4443068869
C   -1.4610751467  -0.6698810397  -0.3475963362
C   -2.7545044961  -1.0588473366  -0.6111703056
C   -3.8532834766  -0.3530728063  -0.0959575495
C   -3.5709509552   0.7816539540   0.6888944778
C   -2.2816352477   1.1688337454   0.9508128590
C   -5.3986414971  -1.7760838726  -1.3343791212
C   -6.2417786670   0.1369826549  -0.0128608804
N    0.0665689379   0.8755570665   0.7270639502
N   -5.1521030842  -0.7595272927  -0.3346543646
O    5.0506712453  -0.7651507297  -0.4665320032
H    3.7220166768   0.9171164039  -1.9216152278
H    1.4305322408   1.6506299008  -1.3744898592
H    1.4604237163  -0.8260692785   2.1551090103
H    3.7520371894  -1.5718522368   1.6253612807
H   -0.6467220747  -1.2379575505  -0.7596041742
H   -2.0922947162   2.0374877407   1.5537528540
H   -4.3674157644   1.3681811276   1.1003313452
H   -2.9033897756  -1.9235950619  -1.2259169393
H   -7.1770602161  -0.3396712327  -0.2668594463
H   -6.2678818313   0.3510628229   1.0488157347
H   -6.1829002504   1.0814183074  -0.5521309054
H   -6.4592073575  -1.9747838197  -1.3816213307
H   -5.0621371970  -1.4805991201  -2.3273390399
H   -4.9085161839  -2.7054985262  -1.0704710838
```
Formaldehyde at the FC geometry
```
O     0.076532925504   -0.000007308878   -2.188426056577
C     0.076533647358    0.000000130593   -0.927928685965
H     0.076536297554    0.929732476449   -0.348240200863
H     0.076536297583   -0.929725299162   -0.348229251592
```
Formaldehyde at the CI geometry
```
O     0.076564820215    0.000020684523   -2.723172193563
C     0.076589440646    0.000006202916   -0.545888753973
H     0.076492452646    1.046774857455   -0.271925067765
H     0.076492454496   -1.046801745893   -0.271838179700
```
*Trans*-butadiene at the FC geometry
```
C    -2.121851103927   -0.013970366404   -2.819555932616
H    -2.122839178616    0.922894141127   -2.279898671033
H    -2.123404142309   -0.914818093581   -2.222478212556
C    -2.118644218820   -0.056097902206   -4.167537566046
H    -2.117671159696   -1.012134976048   -4.678008836243
C    -2.116332323059    1.153623844734   -5.031565880622
H    -2.117344343260    2.109686284849   -4.521131488981
C    -2.113235341454    1.111512834957   -6.379545859464
```

| | | | |
|---|---|---|---|
| H | -2.111579972655 | 2.012409752468 | -6.976570011482 |
| H | -2.112139145209 | 0.174691543100 | -6.919271132961 |